\documentclass[aps,pra,reprint,unsortedaddress,amsmath,amssymb]{revtex4-1}

\usepackage{graphicx}
\usepackage[colorlinks]{hyperref}

\begin{document}


\title{Electron Paramagnetic Resonance Spectroscopy of Er$^{3+}$:Y$_2$SiO$_5$ using a Josephson Bifurcation Amplifier: Observation of Hyperfine and Quadrupole Structures}


\author{Rangga P. Budoyo}
\email{rangga.budoyo@lab.ntt.co.jp}

\author{Kosuke Kakuyanagi}
\author{Hiraku Toida}
\author{Yuichiro Matsuzaki}
\author{William J. Munro}
\author{Hiroshi Yamaguchi}
\author{Shiro Saito}
\affiliation{NTT Basic Research Laboratories, NTT Corporation, 3-1 Morinosato Wakamiya, Atsugi, Kanagawa 243-0198 Japan}


\date{\today}

\begin{abstract}
We performed magnetic field and frequency tunable electron paramagnetic resonance spectroscopy of an Er$^{3+}$ doped Y$_2$SiO$_5$ crystal by observing the change in flux induced on a direct current-superconducting quantum interference device (dc-SQUID) loop of a tunable Josephson bifurcation amplifer. The observed spectra show multiple transitions which agree well with the simulated energy levels, taking into account the hyperfine and quadrupole interactions of $^{167}$Er. The sensing volume is about 0.15~pl, and our inferred measurement sensitivity (limited by external flux noise) is approximately $1.5\times10^4$ electron spins for a 1~s measurement. The sensitivity value is two orders of magnitude better than similar schemes using dc-SQUID switching readout.
\end{abstract}


\maketitle



Electron paramagnetic resonance (EPR) is the gold standard tool to characterize the spin and magnetic properties of a wide range of materials~\cite{abragam_epr1970, scheweiger_pepr2001}. Over several decades, there has been numerous efforts to improve the sensitivity of EPR measurements, including magnetic resonance force microscopy~\cite{rugar_nature2004a, mamin_natnano2007a} and using a single NV center in diamond~\cite{maze_nature2008,balasubramanian_nature2008a}. In recent years, significant improvements were achieved by operating in cryogenic temperatures, where thermal noise is significantly reduced and the spins are highly polarized. In such cases the spins of interest are coupled to superconducting planar resonators~\cite{schuster_prl2010a, kubo_prl2010a} or 3D cavities~\cite{probst_prb2014a}, resulting in avoided crossing or reduction of resonator quality factor.
{By combining superconducting lumped-element resonators with small magnetic mode volume and quantum-limited parametric amplifiers, recently sensitivity values of pulsed EPR have reached $\sim60$~spins$/\sqrt{\text{Hz}}$~\cite{probst_apl2017a}.} 
However, these EPR schemes typically only allow magnetic field-dependent study, because the operating frequency range is limited to frequencies near the resonator's resonance. 
{Frequency and magnetic field-dependent EPR spctroscopy can be advantageous when compared to conventional, single-frequency EPR, in the characterization of more complicated materials~\cite{vanslageren_pccp2003a, wiemann_apl2015a, chen_arxiv2017a}, \textit{e.g.}, anisotropic materials, systems with electron spin $S>1/2$ (zero-field interactions), or systems with nuclear spin $I>0$ (hyperfine and quadrupole interactions).}
Efforts to extend the frequency range include using a tunable resonator~\cite{kubo_prl2010a, chen_prb2016a, chen_arxiv2017a}, a resonator with multiple narrowly spaced modes (\textit{e.g.}, a whispering gallery mode resonator~\cite{benmessai_prb2013a, farr_prb2013a}), or a broadband waveguide~\cite{clauss_apl2013a, wiemann_apl2015a}. However, these schemes typically suffer from worse sensitivity or still limited frequency range. 

EPR can also be performed by measuring the magnetization from the polarization of spins using a sensitive magnetometer. This method allows both magnetic field and frequency-dependent measurements. One typical magnetometer is a direct current-superconducting quantum interference device (dc-SQUID)~\cite{chamberlin_jlowtempphy1979a, cage_rsi2004a, cage_jap2005a}, with continuous wave (cw)-EPR sensitivities reaching $\sim10^6$~spins$/\sqrt{\text{Hz}}$~\cite{toida_apl2016a}. The stated sensitivity was limited by measurements using switching readout to obtain the critical current of the SQUID, which involves switching to voltage state and causes heating. Heating necessitates a significant relaxation time after a single measurement and thus limits repetition rate and sensitivity~\footnote{We note that dc-SQUID magnetometers are often operated not using switching measurements but in continuous mode where the SQUID is always in voltage state~\cite{clarke_squid2006}.}.

In this paper, we introduce a method to perform cw-EPR spectroscopy using a Josephson bifurcation amplifier (JBA) as a magnetometer, with more than two orders of magnitude better sensitivity compared to EPR using dc-SQUIDs. A JBA consists of a superconducting resonator with a Josephson junction as a nonlinear element~\cite{siddiqi_prl2004a, siddiqi_prl2005a}. JBAs are typically used for readout of superconducting quantum bits (qubits)~\cite{siddiqi_prb2006a, lupascu_prl2006a}. In a JBA magnetometer, the junction is replaced by a dc-SQUID, thus the resonance is tunable by changing the flux penetrating the dc-SQUID~\cite{vijay_rsi2009a}. The EPR sensitivity improvement was due to the increased detection sensitivity from operating in the bifurcation regime, as well as higher measurement repetition rates possible compared to switching measurements using dc-SQUIDs~\cite{toida_apl2016a}. We used the JBA to perform cw-EPR on Er$^{3+}$ doped Y$_2$SiO$_5$ (YSO) crystal. Recently there has been increased interest in using Er$^{3+}$ ions in solids as qubits due to their long coherence times and the presence of optical transition at 1.5~$\mu$m telecom $C$ band~\cite{bottger_prb2006a}. At millikelvin temperatures, coupling has been observed between Er dopants in various crystals and resonators~\cite{probst_prb2014a, bushev_prb2011a, probst_prl2013a, probst_apl2014a, probst_prb2015a}.


\begin{figure*}[tpb]
\includegraphics[width=1.6\columnwidth]{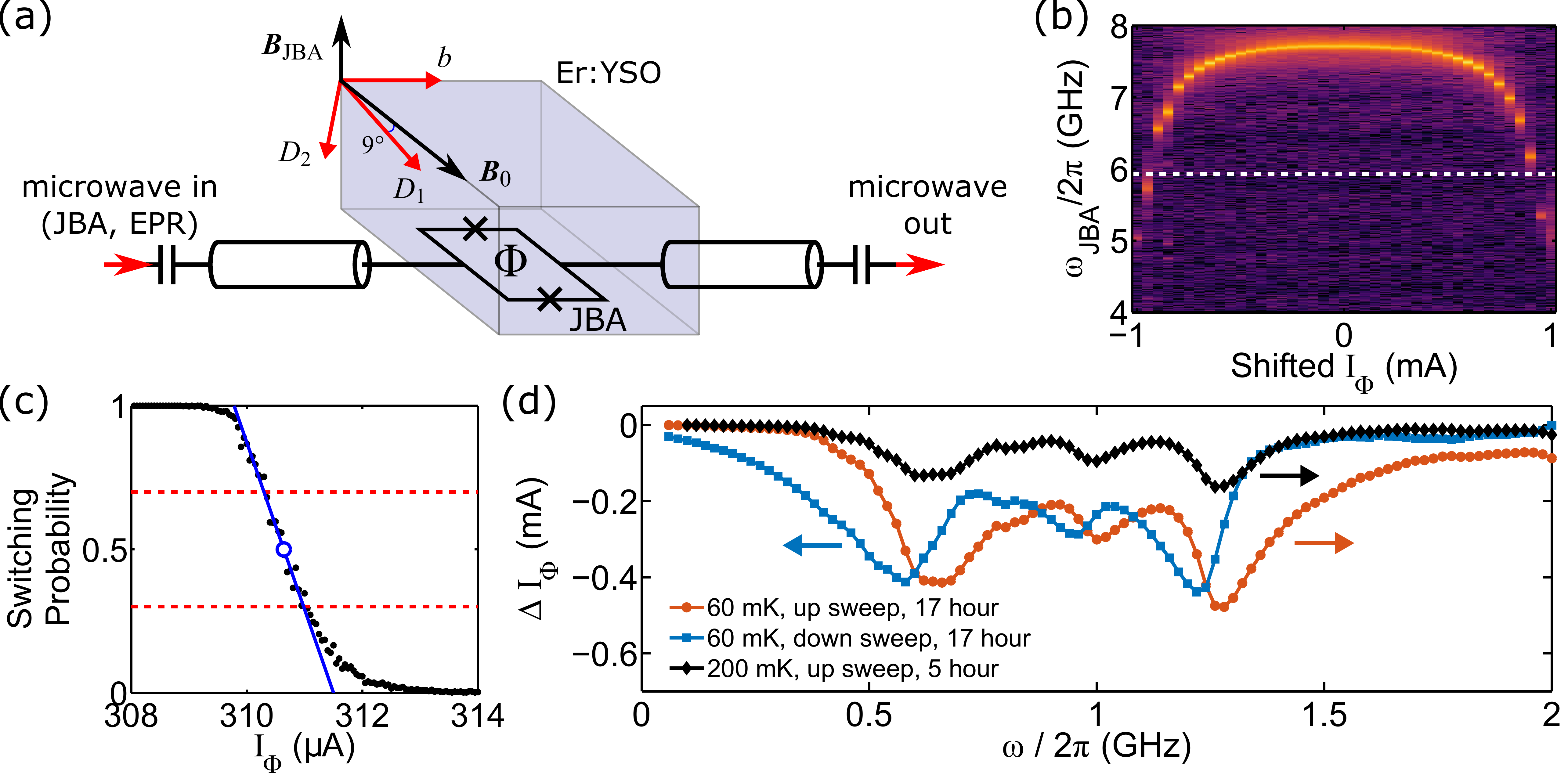}%
\caption{\label{fig:Fig1_combined}(Color online) (a)~Schematic diagram of EPR spectroscopy of Er:YSO using JBA. (b)~Color log plot of transmission $|S_{21}|$ of JBA resonance without YSO crystal as a function of flux bias current $I_\Phi$ over one flux quantum $\Phi_0$. $I_\Phi$ value is shifted such that $I_\Phi=0$ corresponds to zero flux penetrating the dc-SQUID. The dashed line at 5.93~GHz marks the JBA frequency chosen for EPR spectroscopy. (c)~Switching probability of the JBA as a function of $I_\Phi$. The dashed lines at 30\% and 70\% define the boundary for linear fit (solid line) to determine $I_\Phi$ value for 50\% (open circle). (d)~Comparison of EPR spectra for $B_0=5.42$~mT at 60~mK and~200 mK. The red (blue) curve was taken at 60~mK for a 17-hour long measurement with increasing (decreasing) frequency sweep. The asymmetry and the hysteresis of the structures can be clearly seen. The black curve was taken at 200~mK for a 5-hour long measurement with increasing frequency sweep. The spectrum show reduced asymmetry.}
\end{figure*}

Our experiment was performed inside a dilution refrigerator with a simplified schematic shown in Fig.~\ref{fig:Fig1_combined}(a). The JBA used was a coplanar waveguide resonator with a 12 $\times$ 13~$\mu$m$^2$ dc-SQUID embedded in the center of the resonator, and the resonance frequency was observed between approximately 5.0 and 7.7~GHz depending on the flux $\Phi$ penetrating the dc-SQUID [see Fig.~\ref{fig:Fig1_combined}(b)]. A 200~ppm Er$^{3+}$ doped YSO crystal (Scientific Materials, Inc.) was bonded on top of the dc-SQUID. Two separate superconducting magnets were used to apply magnetic fields $\boldsymbol{B}_0$ and $\boldsymbol{B}_\text{JBA}$. 
$\boldsymbol{B}_0$ was oriented parallel to the chip plane and, according to subsequent electron backscatter diffraction measurements, parallel to the $D_1$-$D_2$ plane of the crystal,  $9^\circ$ from the $D_1$ axis (see Supplemental Material~\cite{suppmat} for details). A much smaller $\boldsymbol{B}_\text{JBA}$ to tune the JBA frequency was oriented perpendicular to the chip plane and parallel to the $D_1$-$D_2$ plane.

The microwave input line (used for both JBA and EPR) was capacitively coupled to the JBA. The JBA was operated in the bistable regime with frequency $\approx5.93$~GHz chosen for both strong bifurcation and ability to detect small changes in flux [see Fig.~\ref{fig:Fig1_combined}(b)]. 
To measure the flux induced by the spin ensemble, we swept $B_\text{JBA}$ by changing the flux bias current $I_\Phi$ to find the flux $\Phi$ corresponding to the bifurcation regime. We employed a pulse sequence similar to that used for qubit readout~\cite{siddiqi_prb2006a, lupascu_prl2006a}, with JBA pulse length of about 2 $\mu$s. For each $I_\Phi$, we averaged over 2500 repetitions with a 10~$\mu$s repetition time. The resulting switching probability (to the high amplitude state of the JBA) as a function of $I_\Phi$ rapidly changed between 0 and 1 at the bifurcation regime. We fit the region where the probability is between 30\% and 70\% to a line  [see Fig.~\ref{fig:Fig1_combined}(c)], and used $I_\Phi$ value for 50\% probability.
For a fixed $B_0$, we swept the EPR microwave frequency $\omega$ and measured the shift $\Delta I_\Phi$ compared to when the microwave was turned off.  

Near the 20~mK  base temperature of the refrigerator, the measured spectra showed asymmetric structures and the shape depended on the frequency sweep direction [see Fig.~\ref{fig:Fig1_combined}(d)]. The asymmetry are most likely due to long Er:YSO spin relaxation times at low temperatures and low magnetic fields~\cite{hastingssimon_prb2008a, probst_prl2013a, probst_apl2014a}. As a result, most of the measurements were done at higher temperatures, where the asymmetry was reduced. We performed EPR spectroscopy at 200~mK for $B_0$ between 0.27 and 6.50~mT and $\omega/2\pi$ between 0.1 and 5.2~GHz. The JBA switching probability curve exhibited distortion for $\omega/2\pi>$ 5.2~GHz as the frequency becomes too close to the JBA frequency, which sets the measurement upper limit of $\omega$~\footnote{EPR can be performed further above the JBA frequency up to the frequency upper limit of the microwave input line, which is about 26~GHz in our measurement setup. We however did not explore this frequency region. EPR may also be performed near the JBA frequency with the JBA in the monostable regime as a tunable resonator.}.


\begin{figure*}[tpb]
\includegraphics[width=1.75\columnwidth]{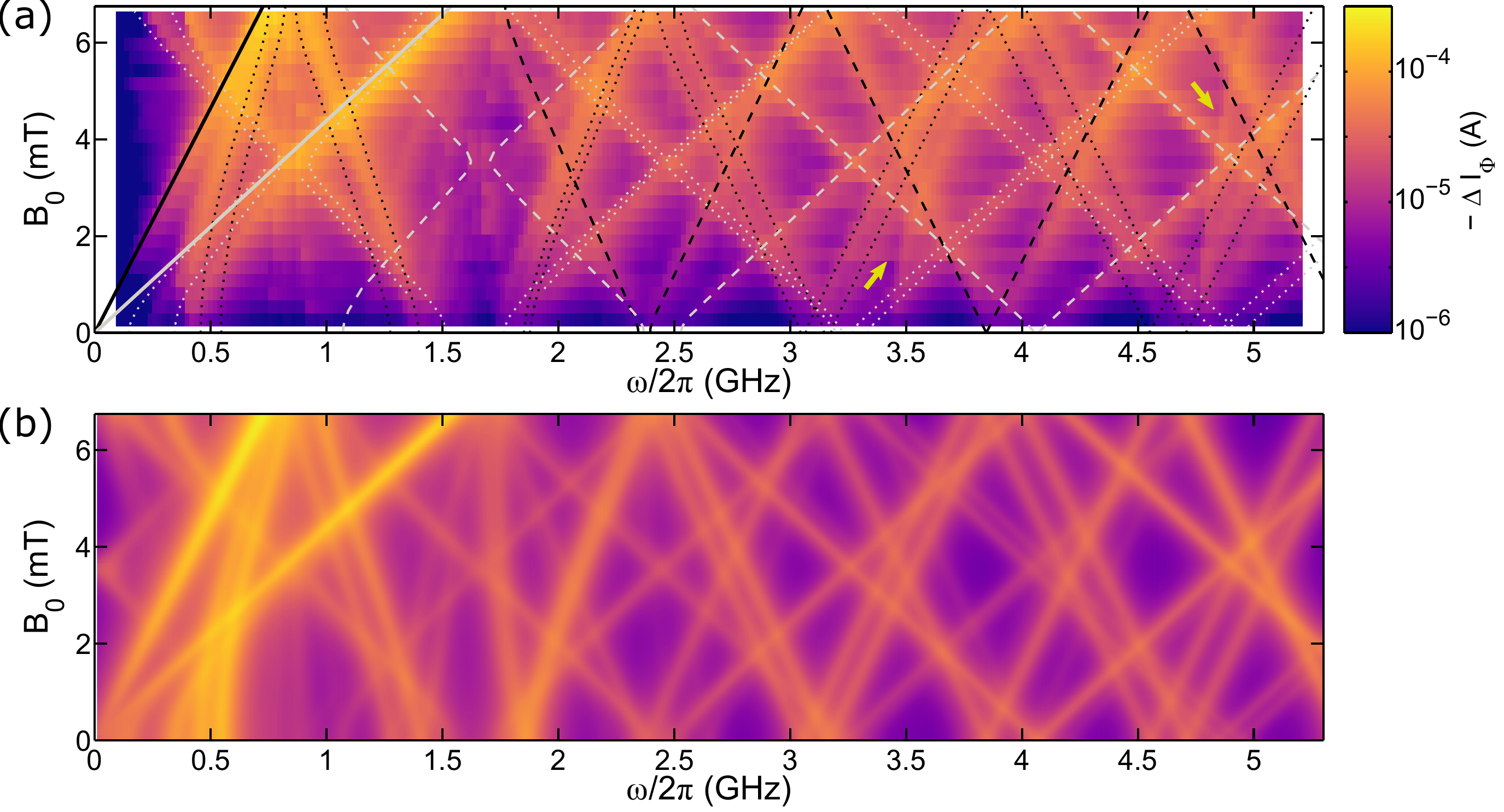}%
\caption{\label{fig:ESR_colorplot}(Color online) (a)~Log plot of measured EPR spectra of Er:YSO at 200~mK. Solid lines mark simulated $I=0$ Er isotope transition frequencies and dashed (dotted) lines mark some simulated $I=7/2$ Er isotope transition frequencies with $\Delta m_I=0$ ($\Delta m_I=\pm 1$), respectively. For these Er transitions, black (grey) lines mark site 1 (2) transitions, respectively. Yellow arrows mark additional transitions due to impurities. (b)~Simulated EPR spectra for the same parameter range.}
\end{figure*}


Figure~\ref{fig:ESR_colorplot}(a) shows the measured spectra, where we observed multiple transitions. We can compare them with the expected spectra given by the usual spin Hamiltonian~\cite{abragam_epr1970}
\begin{equation}
{\cal H} \, = \, \mu_B \boldsymbol{B}_0\cdot\boldsymbol{g}\cdot\boldsymbol{S} +   \boldsymbol{I}\cdot\boldsymbol{A}\cdot\boldsymbol{S} +   \boldsymbol{I}\cdot\boldsymbol{Q}\cdot\boldsymbol{I} - \mu_n g_n   \boldsymbol{B}_0\cdot\boldsymbol{I},
\label{eq:spin_hamiltonian}
\end{equation}
where $\mu_B$ ($\mu_n$)  are the Bohr (nuclear) magneton, $\boldsymbol{S}$ is the electron spin operator, $\boldsymbol{I}$ is the nuclear spin operator, $\boldsymbol{g}$ is the electron g-factor tensor, $\boldsymbol{A}$ is the hyperfine tensor, $\boldsymbol{Q}$ is the quadrupole tensor, and $g_n$ is the nuclear g-factor. In a YSO crystal, the Er$^{3+}$ dopants ($S=1/2$) lie on two crystallographic sites (`site 1' and `site 2') with different sets of $\boldsymbol{g}$, $\boldsymbol{A}$, $\boldsymbol{Q}$, which are all anisotropic. Each site also contains two magnetic inequivalent subclasses related by $C_2$ rotation. However, in our experiment the two subclasses should be equivalent, which is the case when $\boldsymbol{B}_0$ is parallel to $D_1$-$D_2$ plane or $b$ axis of the crystal~\cite{guillotnoel_prb2006a, sun_prb2008a}.

Even isotopes of Er have 77\% natural abundance and $I=0$. Thus for the even isotopes, only the first term on the right side of Eq.~\eqref{eq:spin_hamiltonian} remains, and the transition frequency is given by $\hbar\omega = \mu_B g B_0$, where $g$ is the effective g-factor associated with the direction of $\boldsymbol{B}_0$. Two of the strongest peaks observed [solid lines in Fig.~\ref{fig:ESR_colorplot}(a)] show $\omega\to0$ when $B_0\to0$, with a linear dependence of $B_0$, and thus can be easily identified as the even isotope transitions from the two sites. By fitting the peak frequencies as a function of $B_0$ we extract $g=(7.8\pm0.1)$ and $(16.2\pm0.2)$ for sites 1 and 2, respectively. 
These values are slightly higher than the expected {$g\approx$ 7.0 and 15} for the direction of $\boldsymbol{B}_0$~\cite{guillotnoel_prb2006a, sun_prb2008a, chen_arxiv2017a}. These can be attributed to {$\approx10$\%} enhancement of the magnetic field near the superconducting film due to Meissner effect, comparable to previous reports on superconducting waveguides~\cite{clauss_apl2013a}.

The remaining 23\% abundance consists of $^{167}$Er isotopes with $I=7/2$, with $2\times8=16$ total states. The {majority of the remaining} transitions correspond to the hyperfine transitions between these levels. Some of them appear as two narrowly spaced ($\approx100$~MHz) {nearly} parallel lines, which are characteristics of quadrupole interactions and have been observed in Er:YSO EPR spectra before at higher magnetic fields~\cite{guillotnoel_prb2006a}.
We simulated the expected { transition frequencies and} spectra using the EASYSPIN package~\cite{stoll_jmr2006a}.
We considered all Er sites and subclasses, used the { most recently reported $\boldsymbol{g}$, $\boldsymbol{A}$, and $\boldsymbol{Q}$ values from EPR using a tunable cavity~\cite{chen_arxiv2017a}, and included $B_0$ enhancement effects discussed previously.}
Lines in Fig.~\ref{fig:ESR_colorplot}(a) mark some of the simulated transition frequencies, including allowed hyperfine transitions ($|\Delta m_S|=1$, $\Delta m_I=0$) and dipole-forbidden quadrupole transitions ($|\Delta m_S|=1$, $|\Delta m_I|=1$), where $m_S$ ($m_I$) are the electron (nuclear) spin projection quantum number. The $|\Delta m_I|=1$ transitions are separated into pairs due to the quadrupole interaction as discussed above.
{ At low frequencies, the simulation shows several anticrossing (\textit{e.g.}, at $B_0\approx4$~mT near 0.8 and 1.7~GHz) due to mixing between $|m_S,m_I\rangle$ states, while at higher frequencies the mixing is significantly weaker (see also Supplemental Material~\cite{suppmat}). The simulations also reveal additional hyperfine transitions (\textit{e.g.}, $|\Delta m_I|=2$ and higher). However, some of the observed transtitions (\textit{e.g.} those marked by yellow arrows) do not correspond to Er transitions and can instead be attributed to additional impurities in the YSO crystal~\cite{goryachev_apl2015a,chen_arxiv2017a}.}

Figure~\ref{fig:ESR_colorplot}(b) shows the simulated EPR spectra { for a natural abundance of Er isotopes}.
EASYSPIN assumes a uniform ac magnetic field (from EPR drive) direction and strength, however in the JBA the ac field is not uniform and contains many directions. A full model needs to average different simulations over all spin positions and consider how each spin couples to the static and ac magnetic fields and induces flux on the dc-SQUID. Instead we used an approximate model, a sum of three different simulations, each with the ac field along $\boldsymbol{B}_0$, $\boldsymbol{B}_\text{JBA}$, and crystal $b$ axis. {All transitions are assumed to have 50~MHz linewidth, which is comparable to the observed values and previous reports~\cite{bushev_prb2011a, probst_prl2013a, probst_apl2014a}, and results in the anticrossings appearing as crossings in the simulation.
Overall, there is good agreement between measured and simulated spectra. 
The results confirm the improved spin Hamiltonian parameters~\cite{chen_arxiv2017a} compared to parameters obtained using single-frequency conventional EPR~\cite{guillotnoel_prb2006a}, as well as the advantages of field and frequency-dependent EPR for anisotropic multilevel spin systems. 
We also note that our JBA-EPR frequency range is significantly broader than the 3 to 5~GHz tunable cavity used to derive the $\boldsymbol{g}$, $\boldsymbol{A}$, and $\boldsymbol{Q}$ values~\cite{chen_arxiv2017a}; thus we should be able to improve these parameters even further by also performing EPR at different $\boldsymbol{B}_0$ orientations (\textit{e.g.}, by using a vector magnet~\cite{wisby_prappl2016a}). Indeed, while the agreement between simulated and measured transitions for $^{167}$Er isotopes in site 2 is very good overall, we observed small differences ($<100~$MHz) between the simulated and measured $^{167}$Er site 1 transitions below 2~GHz.}



{Next,} in order to estimate the sensing volume, we need to consider that even though the EPR drive should excite spins over the entire length of the coplanar waveguide, only the spins within the dc-SQUID loop area can induce flux on the dc-SQUID. The area is $\approx 150$~$\mu$m$^2$ and the effective sensing thickness can be estimated to be $\sim1$~$\mu$m~\cite{marcos_prl2010a}. This gives a $\sim150$~$\mu$m$^3$ $=0.15$~pl sensing volume, comparable to the sensing volume of EPR using dc-SQUIDs~\cite{toida_apl2016a} and { the smallest sensing volume of EPR using superconducting resonators~\cite{probst_apl2017a}}. The spin ensemble used in the experiment has a concentration of $3.7\times10^{18}$~spins$/$cm$^{3}$, which means $\approx5.5\times10^8$~spins are located within our sensing volume.

The measurement sensitivity depends on the smallest distinguishable JBA flux shift $\Delta\Phi$, and can be determined from the 95\% confidence bound of the extracted fit value for $I_\Phi$ point with 50\% probability [\textit{e.g.}, Fig.~\ref{fig:Fig1_combined}(c)]. The smallest distinguishable flux shift is about 90 $\mu\Phi_0$ for a 1~s measurement time for each $I_\Phi$ value. 
From spin polarization measurements (see Supplemental Material~\cite{suppmat} for details), the saturation flux is estimated to be $\approx 3.4\Phi_0$. Using this value, we infer a measurement sensitivity of $\approx 1.5\times10^4$~spins, within the range of sensitivities of parametric-amplified pulsed EPR using { typical superconducting lumped-element resonators}~\cite{bienfait_natnano2016a, eichler_prl2017a}. The sensitivity is also about two orders of magnitude better than the sensitivity of EPR using dc-SQUIDs~\cite{toida_apl2016a}. We note that the repetition time in our experiment were 20 times shorter than the repetition time in dc-SQUID EPR measurements, and thus we expect a factor of $\sqrt{20}\approx4.5$ improvement in sensitivity from increased repetition rate. Since the two experiments use dc-SQUIDs of comparable sizes, we conclude that the sensitivity improvement were mainly due to the nonlinearity of the JBA.

\begin{figure}[tpb]
\includegraphics[width=0.95\columnwidth]{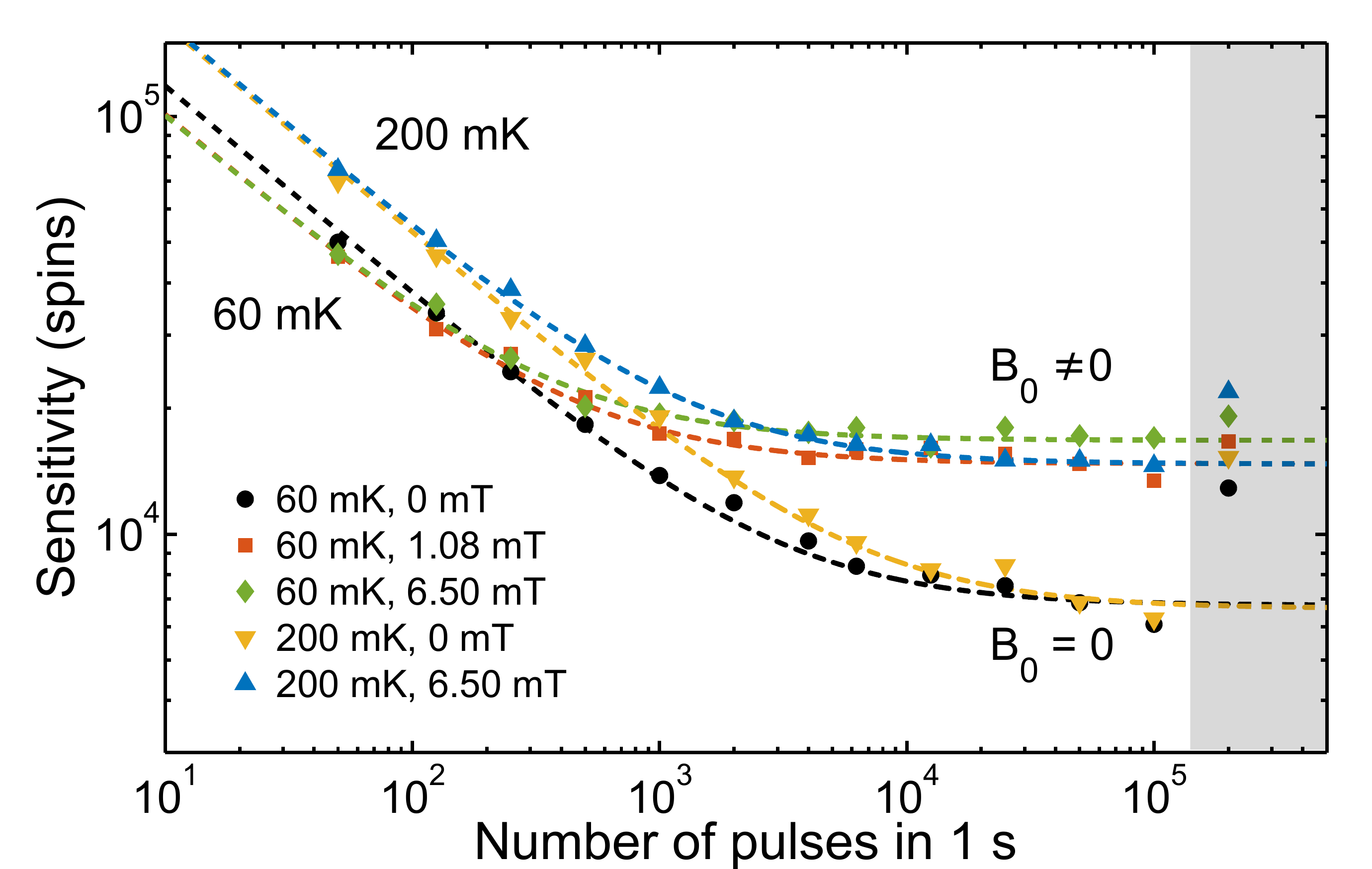}%
\caption{\label{fig:sens_vs_nrep}(Color online) Inferred measurement sensitivity as a function of number of pulses $N$ for several temperature and magnetic field values. Pulse repetition time is varied while the total measurement time for each $I_\Phi$ is kept at 1 second. The dashed curves are the corresponding fit to Eq.~\eqref{eq:sensitivity_fit}. Points in the shaded region are not included in the fit.}
\end{figure}

To understand the limiting source of sensitivity, we inferred the sensitivity from repeated measurements at different $T$ and $B_0$ values. The number of JBA pulses $N$ and the repetition time are also varied, but the product (measurement time) is kept at 1~s. Figure~\ref{fig:sens_vs_nrep} shows the resulting sensitivity plot as a function of $N$. For $N\gtrsim10^5$ (repetition time $\lesssim10$~$\mu$s, shaded region in Fig.~\ref{fig:sens_vs_nrep}), the bifurcation is reduced, resulting in worse sensitivity. This is likely caused by the incomplete JBA relaxation between repetitions, as the pulses are too close together. Otherwise, the sensitivity $S$ for each set of $T$ and $B_0$ values appears to follow the expression
\begin{equation}
S(N) \: = \: \sqrt{\left(\alpha/N\right) + \beta},
\label{eq:sensitivity_fit}
\end{equation}
where $\alpha$ and $\beta$ are fit parameters. $\alpha$ appears to increase with increasing $T$ and independent of $B_0$, and is likely related to the JBA bifurcation profile [Fig.~\ref{fig:Fig1_combined}(c)]. $\beta$, which suggests the presence of coupled external noise, limits the sensitivity at large $N$ to 15000-17000~spins for both measured temperatures and for magnetic fields between 1.08 and 6.50~mT. However, when $B_0=0$, the sensitivity improves to $\approx 7000$~spins, independent of $T$. This suggests when $B_0\neq0$, the sensitivity is  limited by flux noise generated by $B_0$ magnet. Considering this limiting source as well as the configuration of the system, the sensitivity at $B_0=0$ is most likely limited by other external sources, including flux noise from $B_{JBA}$ magnet, flux noise from the environment, as well as mechanical vibrations of the dilution refrigerator. Noise reductions from these sources should allow further sensitivity improvements, approaching the intrinsic JBA sensitivity [$\alpha$ term in Eq.~\eqref{eq:sensitivity_fit}] of $\approx800$~spins$/\sqrt{\text{Hz}}$ for 10~$\mu$s repetition time at 60~mK, and possibly even lower at lower temperatures.


In conclusion, we have demonstrated a scheme to perform cw-EPR spectroscopy over a wide range of magnetic fields and frequencies by observing the change in flux induced on the dc-SQUID of a tunable JBA. We used the scheme on an Er$^{3+}$ doped YSO crystal and observed multiple transitions within the range of measurement, {in good agreement} with spin Hamiltonian simulations including hyperfine and quadrupole interactions of $^{167}$Er isotopes. 
By operating in the bifurcation regime, combined with high repetition rates possible with a JBA, the measurement sensitivity improved by two orders of magnitude when compared to switching measurements of dc-SQUIDs~\cite{toida_apl2016a}, and is comparable to sensitivity of superconducting resonators. The sensitivity can be further improved by reduction of noise from external sources, as well as increase in nonlinearity from refinements in JBA design and operating point. The small sensing volume can also be further reduced, thus improving measurement sensitivity and spatial resolution, by using smaller dc-SQUIDs (\textit{e.g.}, nanoscale SQUIDs~\cite{lam_apl2003a, troeman_nanolett2007a, vasyukov_natnano2013a, martinezperez_acsnano2016a}).

The authors thank  I. Mahboob, K. Shimizu, and { J. Longdell} for valuable discussions. We also thank the staff at NTT Advanced Technology for performing electron backscatter diffraction measurements of the crystal.  This work was supported in part by MEXT Grant-in-Aid for Scientific Research on Innovative Areas ``Science of hybrid quantum systems'' (Grant No. 15H05869 and 15H05870).


\clearpage
\onecolumngrid
\begin{center}
\textbf{\large Supplementary Materials for ``Electron Paramagnetic Resonance Spectroscopy of Er$^{3+}$:Y$_2$SiO$_5$ Using Josephson Bifurcation Amplifier: Observation of Hyperfine and Quadrupole Structures"}

\vskip 1\baselineskip

\parbox{14cm}{\small \hspace{3mm}This supplemental material includes several discussions not included in the main text, which include the determination of magnetic field orientation relative to Er:YSO crystal's optical axes, calculations of the energy levels of the spin system, detection of spin polarization, and measurements of system stability.}

\vskip 1\baselineskip

\end{center}

\twocolumngrid
\setcounter{equation}{0}
\setcounter{figure}{0}
\setcounter{table}{0}
\makeatletter
\renewcommand{\theequation}{S\arabic{equation}}
\renewcommand{\thefigure}{S\arabic{figure}}

\section{\label{sec:axes}Crystal And Magnetic Field Orientation}

To determine the orientation of the optical axes of the Er:YSO crystal, NTT Advanced Technology company performed electron backscatter diffraction (EBSD) measurements of an Er:YSO crystal from the same batch with the same dimensions and the same nominal (vendor-specified) cut as the crystal used in the main text. EBSD can be used to determine the structure, and thus the orientation of the crystallographic axes, of a crystal. 

\begin{figure*}[tpb]
\includegraphics[width=1.5\columnwidth]{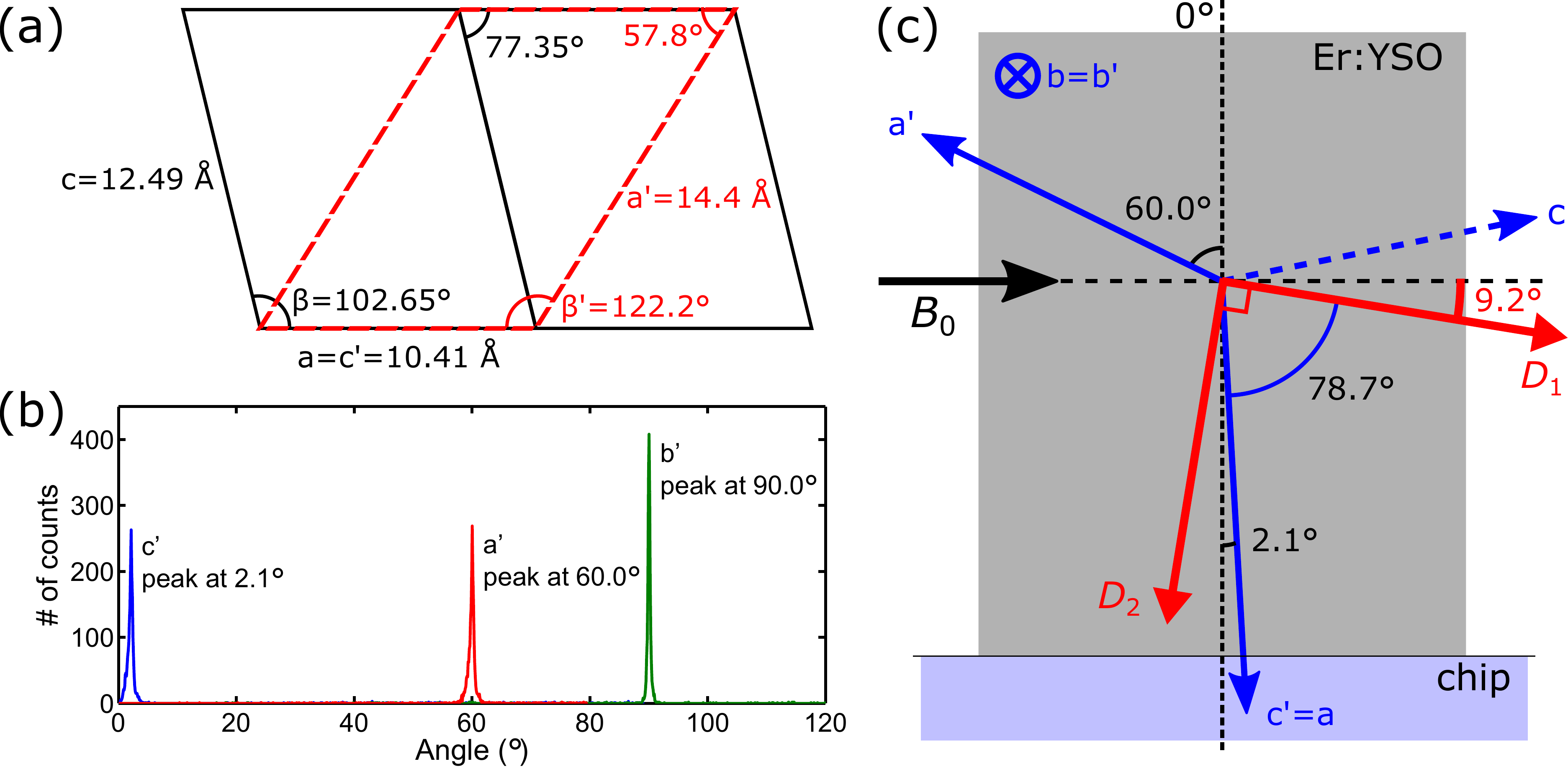}%
\caption{\label{fig:ebsd_summ} (a)~Comparison between the unit cell definitions typically used for YSO crystals (black solid lines) and the one used in EBSD analysis (red dashed lines). $b$ and $b'$ axes are identical and perpendicular to figure plane. 
(b)~Histogram showing inferred orientation of Er:YSO crystallographic axes from analysis of EBSD pattern.
(c)~Alignment of magnetic field $\boldsymbol{B}_0$ and crystal axes used in EPR spectroscopy. Blue arrows denote the directions of the crystallographic axes as determined by EBSD in (b), with 0$^\circ$ line vertical. Red arrows denote the inferred directions of the optical extinction axes.}
\end{figure*}

YSO is a monoclinic crystal with the unit cell typically defined with lattice constants $a=10.41$~\r{A},  $b=6.72$~\r{A},  $c=12.49$~\r{A}, and $\beta=102.65^\circ$~\cite{maksimov_krys1970a}. The analysis of EBSD pattern used an alternate unit cell definition with $a'=14.4$~\r{A},  $b'=b$,  $c'=a$, and $\beta=122.2^\circ$ [Ref.~\cite{michel_cracad1967a}, see Fig.~\ref{fig:ebsd_summ}(a) for comparison between the two definitions], with the results shown in Figs.~\ref{fig:ebsd_summ}(b) and (c).
We found that $a'$-$c'$ plane is parallel to one surface of the crystal [chosen to be the figure plane in Fig.~\ref{fig:ebsd_summ}(c)], with $b'$ perpendicular to this plane as expected. $c'$ is $2.1^\circ$ from the crystal surface which is parallel to chip plane in EPR experiments. Magnetic field $\boldsymbol{B}_0$ was oriented perpendicular to the third crystal surface.
The optical extinction axes $D_1$ and $D_2$ are parallel to $a$-$c$ (or $a'$-$c'$) plane, with $D_1$ axis $78.7^\circ$ from $a$, $23.8^\circ$ from $c$, and perpendicular to $D_2$~\cite{li_ieeejqe1992a}. 
Using these values, we found that $\boldsymbol{B}_0$ was oriented at an angle of about $9.2^\circ$ relative to $D_1$ axis.

The agreement between the observed Er spin frequencies and the simulated frequencies using this specific direction of $\boldsymbol{B}_0$, especially for even Er spins, shows very good alignment of the magnetic field and the crystal within our measurement setup [see Fig.~1(a) of main text]. In addition, using the SQUID JBA as a magnetometer, we have verified that $\boldsymbol{B}_0$ is very nearly parallel to chip surface, with alignment error less than $0.5^\circ$.

\section{\label{sec:spinpol}Energy Levels of the $I=7/2$ Spin System}

\begin{figure*}[tpb]
\includegraphics[width=1.5\columnwidth]{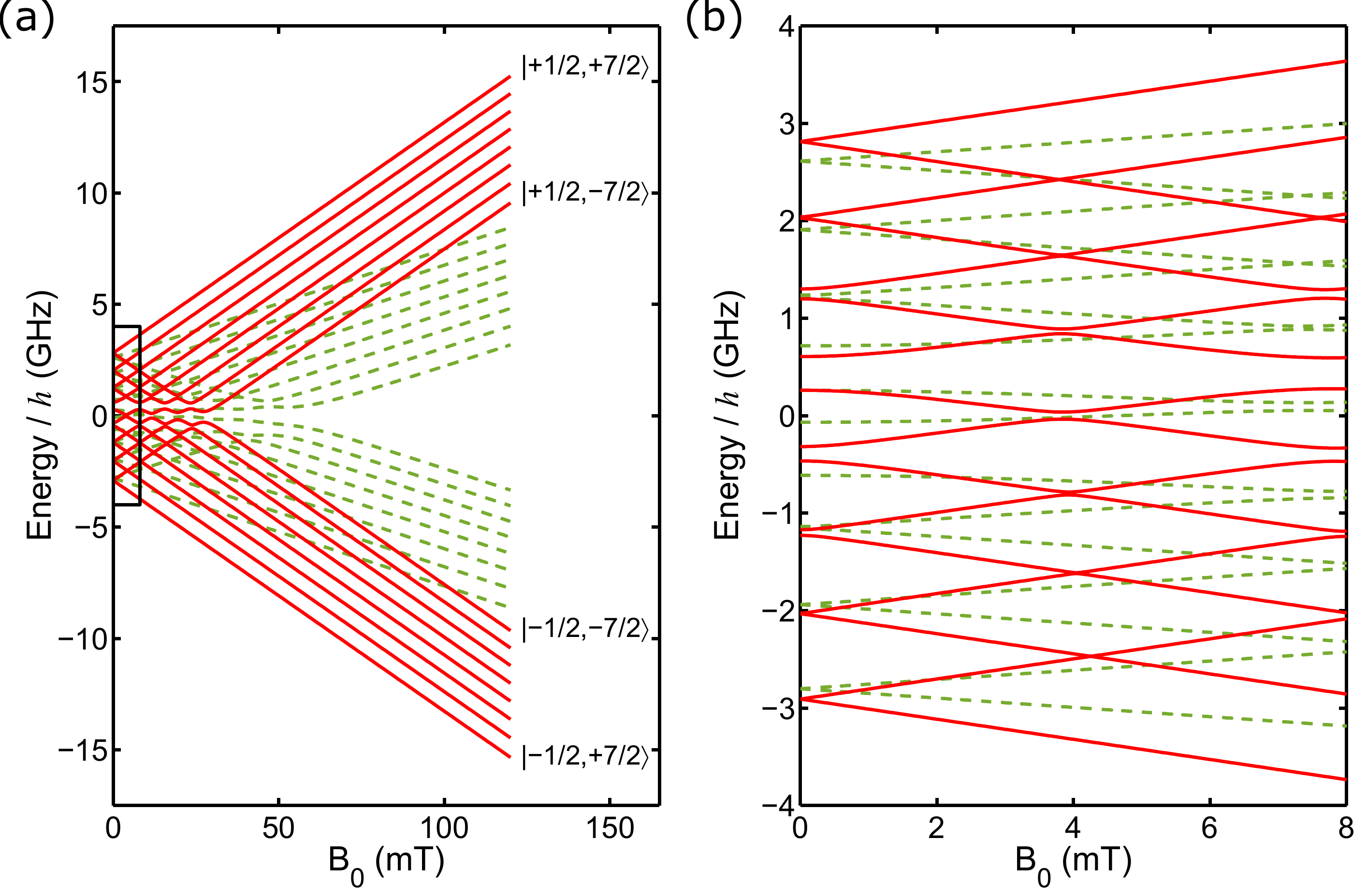}%
\caption{\label{fig:Elevels_comb} Energy level diagram for $^{167}$Er for both site~1 (green dashed lines) and site~2 (red solid lines) as a function of magnetic field $B_0$. (b)~is a zoomed-in view of the area inside the box in (a), and shows the approximate $B_0$ range used in the experiment.}
\end{figure*}

Figures~\ref{fig:Elevels_comb} show the energy level diagram of $^{167}$Er ($I=7/2$) isotopes for both crystallographic sites as a function of applied magnetic field $B_0$. The energy levels were calculated using EASYSPIN~\cite{stoll_jmr2006a},  the { recently reported values} of $\boldsymbol{g}$, $\boldsymbol{A}$, and $\boldsymbol{Q}$~\cite{chen_arxiv2017a}. At large $B_0$, above approximately 80~mT (50~mT) for site 1 (2) respectively, the energy levels appear as two sets of eight parallel lines [see Fig.~\ref{fig:Elevels_comb}(a)]. Here the system is within the Paschen-Back regime where the hyperfine term in the spin Hamiltonian can be written as $\tilde{A} I_z S_z$ and the states can be written as $|m_S,m_I\rangle$. $\tilde{A}$ is the effective hyperfine constant, $z$-axis is defined as the direction of $\boldsymbol{B}_0$, $I_z$ ($S_z$) are the nuclear (electron) spin operator along $z$-axis, and $m_I$ ($m_S$) are the nuclear (electron) spin projection quantum number along $z$-axis. However, our measurements were performed at very low magnetic field $B_0\le6.50$~mT [see Fig.~\ref{fig:Elevels_comb}(b)]. 
Within this range, hyperfine and quadrupole interactions can cause mixing between states. { Mixing results in the avoided level crossings seen in Fig.~\ref{fig:Elevels_comb}(b) and near these crossings the eigenstates are superpositions of the mixed $|m_S,m_I\rangle$ states. Within the experimental $B_0$ range, $m_I=\pm1/2$ states show the strongest mixing, while $m_I=\pm5/2$ and $\pm7/2$ states show very weak mixing.}

\section{\label{sec:spinpol}Spin Polarization Measurement}

For a simple paramagnetic spin $1/2$ system with $N$ spins at temperature $T$, the total magnetic moment of the spins $M$ under applied magnetic field $B_0$ follows the Langevin paramagnetic equation given by
\begin{equation}
M \: = \: \frac{N\mu_B g}{2}  \, \tanh\left(\frac{\mu_B g B_0}{2 k_B T}\right),
\label{eq:magmom}
\end{equation}
where $\mu_B$ is the Bohr magneton and $g$ is the g-factor. Thus saturation appears when $(\mu_B g B_0/2 k_B T)\to\infty$, with $M\to N\mu_B g/2$. As the induced flux $\Phi$ and thus the flux bias current $I_\Phi$ are proportional to the induced magnetization, $I_\Phi$ depends on $T$ and $B_0$ as
\begin{equation}
I_\Phi \: = \: I_s \, \tanh\left(\frac{\mu_B g B_0}{2 k_B T}\right),
\label{eq:iphi_pol}
\end{equation}
where $I_s$ is the flux bias shift corresponding to saturation. For Er:YSO, where $g$ is anisotropic and there are multiple transitions with different $g$ factors, Eq.~\eqref{eq:iphi_pol} becomes 
\begin{equation}
I_\Phi \: = \: I_s  \, \tanh\left(\frac{\mu_B \tilde{g} B_0}{2 k_B T}\right),
\label{eq:iphi_pol_er}
\end{equation}
where $\tilde{g}$ is the effective g-factor. With the configuration of the YSO crystal and magnetic fields in our experiment, $\tilde{g}$ is estimated, and has been measured in a dc-SQUID magnetometry in a similar configuration, to be approximately 6.2~\cite{toida_apl2016a}. 

\begin{figure}[tpb]
\includegraphics[width=\columnwidth]{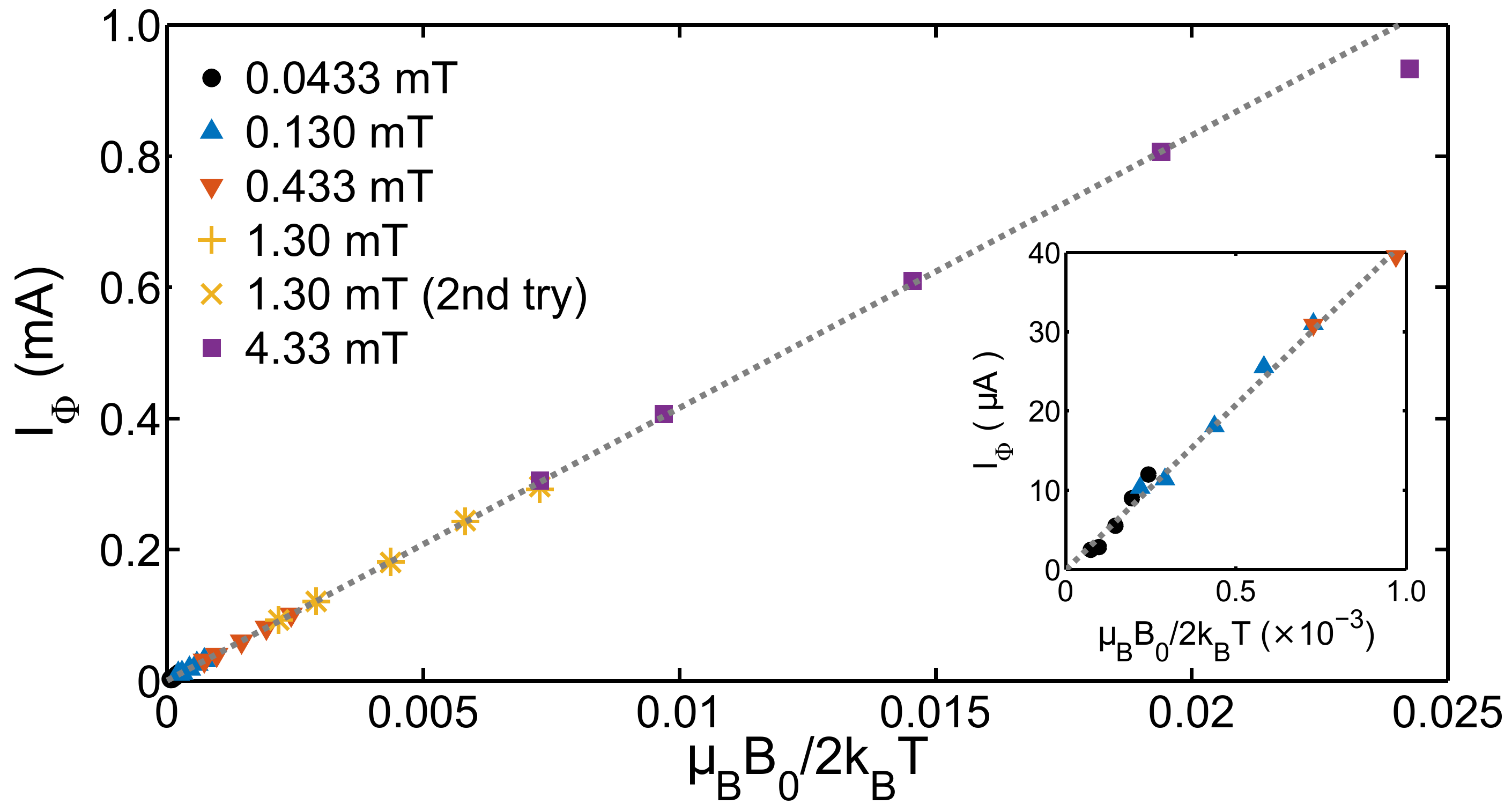}%
\caption{\label{fig:SpinDet} Flux bias shift $I_\Phi$ as a function of $\mu_B B_0/2k_B T$. Different colors show different magnetic field values. The dashed line show a linear fit to data, not including the point at highest $\mu_B B_0/2k_B T\approx 0.024$. The inset shows a zoomed-in view at small shift and $\mu_B B_0/2k_B T$ values. }
\end{figure}

We measured the spin polarization of Er:YSO by changing both $T$ and $B_0$. However, even small misalignment in $B_0$ may introduce additional magnetic field component perpendicular to chip surface and introduce shift in $I_\Phi$. As we discuss in Section~\ref{sec:stability} below, trapped vortices may also introduce additional shift. To take this into account, in the measurement we varied $T$ while keeping $B_0$ constant for each set of measurement. Different sets have different $I_\Phi$ offset, and Fig.~\ref{fig:SpinDet} show the combined $I_\Phi$ data as a function of $\mu_B B_0/2k_B T$ with offset adjustment applied. The data generally follow a linear trend, except at the largest shift value ($\mu_B B_0/2k_B T\approx0.024$). This behavior agrees with  Eq.~\eqref{eq:iphi_pol_er}, as when $(\mu_B g B_0/2 k_B T)\ll1$, Eq.~\eqref{eq:iphi_pol_er} can be approximated to
\begin{equation}
I_\Phi \: = \: I_s  \, \frac{\mu_B \tilde{g} B_0}{2 k_B T}.
\label{eq:iphi_pol_er_cw}
\end{equation}
As we did not observe full saturation, fitting the data to Eq.~\eqref{eq:iphi_pol_er} is difficult. Instead we fit the data within the linear regime to Eq.~\eqref{eq:iphi_pol_er_cw}, where we use $\tilde{g}=6.2$ and $I_s$ is the only fitting parameter. The extracted $I_s\approx6.7$~mA corresponds to a saturation flux shift of $3.4\Phi_0$. This value is consistent within 20~\% with the extracted saturation flux value from the previously discussed dc-SQUID measurements~\cite{toida_apl2016a}, considering the ratio of the dc-SQUID loop areas.

We also note at the lowest $B_0$ value of 43.3~$\mu$T (see inset of Fig.~\ref{fig:SpinDet}), the gradient of the line appeared to be slightly higher than for other $B_0$ values. This can be explained by the presence of residual magnetic field, suggesting a residual field of several $\mu$T at the direction of $B_0$.

\section{\label{sec:stability}JBA Stability}

\begin{figure}[tpb]
\includegraphics[width=\columnwidth]{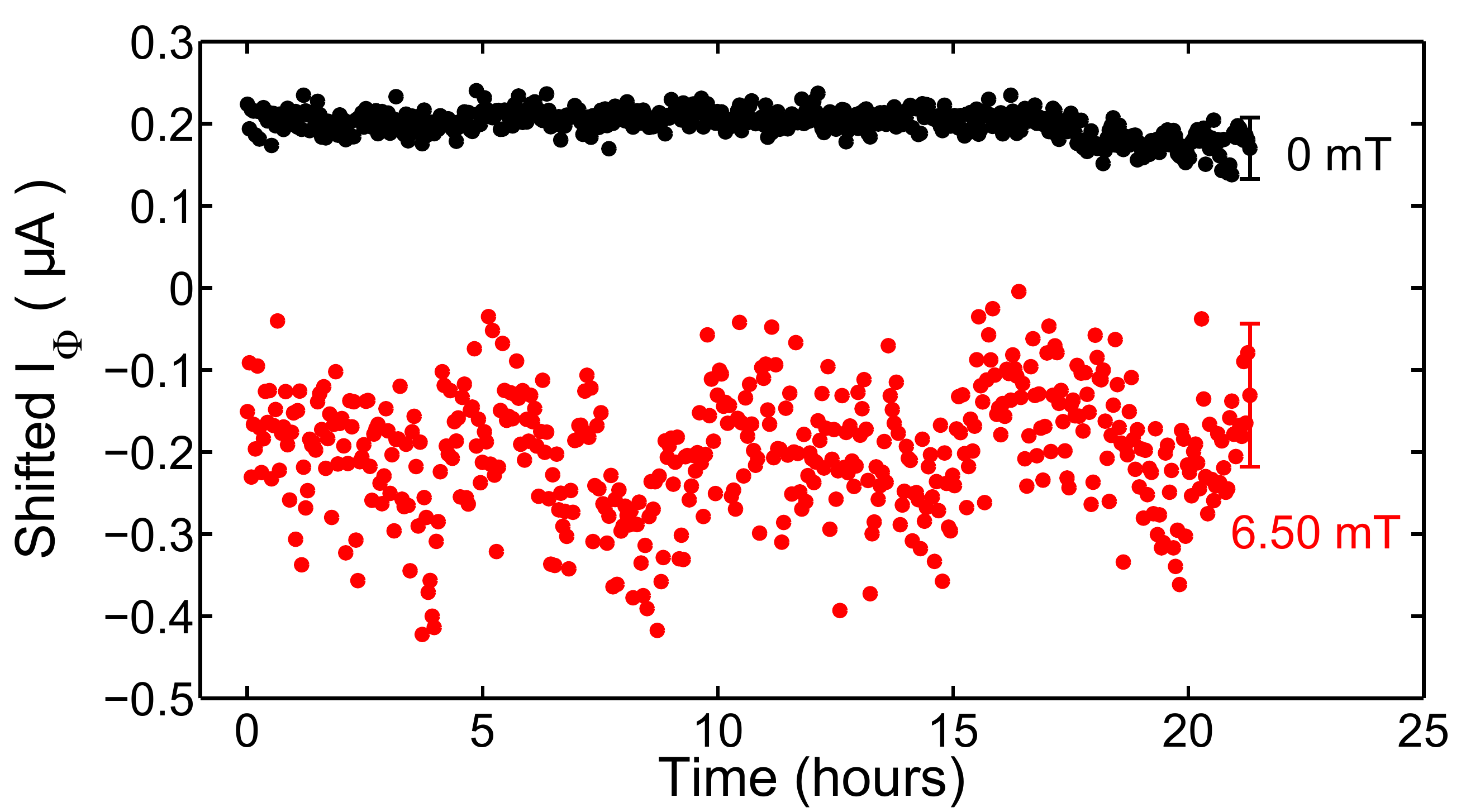}%
\caption{\label{fig:stability_test} Repeated measurements of the JBA flux bias $I_\Phi$ over 21~hours without EPR drive for $B_0=0$ (black dots) and $B_0=6.50$~mT (red dots) at 200~mK. The $I_\Phi$ values are offset such that the average $I_\Phi$ is at 0.2~$\mu$A for $B_0=0$ and at $-0.2$~$\mu$A for $B_0=6.50$~mT. The error bars represent the averaged 95\% confidence bound in determining $I_\Phi$ from the fits of switching probability curves with $10^5$ pulses and 10~$\mu$s repetition time for each $I_\Phi$.}
\end{figure}

To check the stability of the system, we performed repeated measurements of $I_\Phi$ over 21~hours, at $T=200$~mK and for both $B_0=0$ and $B_0=6.50$~mT. During the measurement, no EPR drive was applied. The results are shown in Fig.~\ref{fig:stability_test}, with the absolute $I_\Phi$ shifted for visibility. The error bars represent the averaged 95\% confidence bound in determining $I_\Phi$ from fits of the JBA switching probability curves with $10^5$ pulses and 10~$\mu$s repetition time for each $I_\Phi$. The error bar size for $B_0=0$ is about half of that for $B_0=6.50$~mT. This is consistent with the ratio of sensitivity for the two $B_0$ values discussed in the main text. For $B_0=0$, $I_\Phi$ value appeared to be stable for most of the range of measurement, although a possible slow and small drift was observed around hour 17-18. The source of this apparent drift is unclear, but may be due to changes in residual field.

For $B_0=6.50$~mT, we observed what appeared to be periodic fluctuations with period approximately 5~hours, with peak-to-peak amplitude of approximately 0.2~$\mu$A. This is two to three orders of magnitude less than the typical $I_\Phi$ shift at observed transitions [see Fig.~2(a) of main text]. Thus even though the reference $I_\Phi$ level appeared to fluctuate with time, this effect should be negligible in EPR experiments where we typically set the refererence level from a single measurement. The 5-hour period appeared to be consistent with the observed periodic fluctuations of cooling water temperature for the dilution refrigerator~\footnote{The water temperature fluctuations were also present when the refrigerator was not in operation, which means the source of the fluctuations was external, such as the performance of the water chiller or the overall load of the cooling water system}. This suggests that the fluctuations in $I_\Phi$ were due to fluctuations in temperature of the mixing chamber of the dilution refrigerator. From Eq.~\eqref{eq:iphi_pol_er_cw}, the corresponding bias change $\delta I_\Phi$ due to a small temperature change $\Delta T$ is given by
\begin{equation}
\delta I_\Phi \: \approx \: \frac{I_s\mu_B \tilde{g} B_0}{2 k_B T^2}\,\delta T.
\label{eq:iphi_tfluc}
\end{equation}
$\delta I_\Phi=0.2$~$\mu$A corresponds to temperature change $\delta T\approx0.09$~mK at~200 mK and 6.50~mT. The uncertainty of the thermometer for the mixing chamber is observed to be approximately 0.3-0.4~mK at 200~mK, about three to four times larger than the estimated drift. This again suggests that $I_\Phi$ fluctuations were caused by small refrigerator temperature fluctuations, which were not observed by the thermometer and not corrected by the refrigerator temperature control. We note that this temperature fluctuations should also exist during $B_0=0$ measurements, but as shown in Eq.~\eqref{eq:iphi_tfluc} they do not contribute to $I_\Phi$ fluctuations at $B_0=0$, and contribute trivially for the estimated amount of residual field (see previous section). 

Finally, we also note that vortices trapped on superconducting film also contributed to additional shift in $I_\Phi$. The release and movement of these vortices could result in an instantaneous jump in $I_\Phi$ that is distinguishable from $I_\Phi$ drift due to fluctuations in temperature or changes in residual field. Typically jumps occurred when flux bias was swept too much or too rapidly. To reduce the likelihood of jumps, we generally limited our measurements to a regime where we expect flux shifts $<\Phi_0$. However, in this regime jumps can still rarely occur, necessitating an adjustment of reference $I_\Phi$.

\bibliography{JBAESR}

\end{document}